\documentclass[aps,nofootinbib,superscriptaddress, showpacs,preprintnumbers, nofootinbibt,twocolumn]{revtex4-2}

\usepackage{epsfig}
\usepackage{multirow}
\usepackage{subcaption}
\usepackage{eurosym}
\usepackage{dcolumn}
\usepackage{bm}
\usepackage{enumerate}
\usepackage{float}
\usepackage{epstopdf}
\usepackage{amsmath}
\usepackage{bm}
\usepackage{amsfonts}
\usepackage{amssymb}
\usepackage{graphicx}
\usepackage{alphalph,mathtools}
\usepackage{etoolbox}
\usepackage{color}
\usepackage{booktabs}
\usepackage{hyperref}
\hypersetup{colorlinks,citecolor=blue}
\usepackage{footnote}
\usepackage{makecell,tabularx}

\usepackage{graphicx} 

\def\be{\begin{equation}}
\def\ee{\end{equation}}
\def\bea{\begin{eqnarray}}
\def\eea{\end{eqnarray}}

\begin{document}

\title{Charged Black Hole Solution in Hoyle–Narlikar Gravity and its Thermodynamic Properties} 

\author{Aniruddha Ghosh}
 \email{ruddha.g@gmail.com}
 \affiliation{%
 Department of Mathematics, Indian Institute of Engineering Science and Technology, Shibpur, Howrah-711103, India.\\ 
}%
\author{Ujjal Debnath}%
 \email{ujjaldebnath@gmail.com}
\affiliation{%
 Department of Mathematics, Indian Institute of Engineering Science and Technology, Shibpur, Howrah-711103, India.\\ 
}%

\begin{abstract}
In this paper, we derive a new class of analytic black hole solutions within the framework of Hoyle--Narlikar gravity, where the black hole is surrounded by an electric charge acting as the source. After obtaining the corresponding metric, we investigate the thermodynamic properties of the resulting black hole solutions, including the Hawking temperature and entropy. Subsequently, we compute the specific heat and Gibbs free energy in order to analyse the thermodynamic stability of the black hole in Hoyle--Narlikar gravity. An interesting result of our analysis is that the black hole becomes thermodynamically stable for large values of the horizon radius within this framework.
\par
\vspace{0.1cm}
\end{abstract}

\maketitle
\section{Introduction}\label{sec1}
The discovery in the late 1990s that the Universe is undergoing an accelerated phase of expansion~\cite{I1,I2,I3,I4,I5,capozziello2002curvature,carroll2004cosmic,sotiriou2006f,hu2007models} created a significant challenge for the conventional description provided by General Relativity (GR). 
This surprising observational result stimulated extensive theoretical efforts to explain the mechanism responsible for this cosmic acceleration. 
In general, the proposed explanations can be classified into two main approaches. 
The first approach assumes the presence of additional, non-luminous components within the standard relativistic framework, usually identified as dark matter~\cite{I1.4,I1.5,I1.6,arkani2009theory,bertone2018history,cirelli2026dark} and dark energy~\cite{I1.1,I1.2,I1.3,copeland2006dynamics,li2011dark,frieman2008dark,comelli2003dark}. 
The second approach focuses on extending or modifying the theory of gravity itself by generalising the Einstein--Hilbert action, often through the inclusion of functions involving curvature invariants or matter-related quantities.

In this work, we follow the second approach and study an alternative theory of gravity. 
In particular, we consider the Hoyle--Narlikar model, which represents a modification 
of Einstein’s theory by introducing a creation field that changes the form of the 
gravitational field equations. This model offers a useful framework for studying 
the large-scale evolution of the Universe as well as the properties of compact 
astrophysical objects such as black holes.
\par
The creation field theory, first proposed by Hoyle and later developed together 
with Narlikar~\cite{I6,I7,I8}, was introduced as an alternative cosmological model. 
In this theory, the Universe expands continuously while the overall matter density 
remains constant. To achieve this, Einstein’s field equations are modified by 
including a creation field. This field possesses negative energy and allows matter 
to be created continuously as the Universe expands. As a result, the model provides 
a natural way to maintain a constant matter density and also offers possible 
solutions to the horizon and flatness problems that arise in the standard Big Bang 
cosmology.
\par
Narlikar~\cite{I9} later showed that the creation of matter in this theory can be 
understood as a process that occurs at the expense of a negative-energy creation 
field. Subsequently, Narlikar and Padmanabhan~\cite{I10} pointed out that Einstein’s 
field equations allow solutions in which radiation can be represented by a 
massless and chargeless scalar field $C$ carrying negative energy. Hawking~\cite{I11} 
also examined this idea by studying the implications of negative mass within the 
framework of Hoyle--Narlikar gravity. In this context, Hoyle~\cite{I12} argued that 
the continuous production of matter can be incorporated consistently into the 
structure of General Relativity, leading to a Universe that expands steadily while 
remaining dynamically stable without the need for a cosmological constant.
\par
Over the years, several researchers have explored different features of the 
Hoyle--Narlikar theory in relation to cosmology as well as particle physics. 
McIntosh~\cite{I13} extended the particle theory proposed by Hoyle and Narlikar by 
considering the implications of Mach’s principle. In addition, Davies~\cite{I14} 
showed that this gravitational theory creates a direct connection between 
cosmological quantities and the properties of elementary particles. Later, 
Hoyle and Narlikar~\cite{I15} investigated the conformal structure of the theory, 
revealing some of its distinctive geometric characteristics.
\par
Later investigations~\cite{I16} explored different cosmological models within this 
framework, including both static Universe models and those based on the Big Bang 
scenario. These studies also discussed the observational importance of key 
cosmological parameters such as the Hubble constant and the cosmological constant. 
Thorne~\cite{I17} studied several consequences of the theory, including its 
possible effects on the formation of primordial elements, the origin of early 
cosmic magnetic fields, and the large-scale isotropy observed in the Universe. 
In addition, the role of dark energy and the accelerated expansion of the Universe 
have been examined within the Hoyle--Narlikar cosmological model~\cite{I18}.

In this paper, we construct black hole solutions within the framework of Hoyle--Narlikar gravity, where the source is taken to be an electric charge. We first derive an approximate solution for the black hole metric in this framework. After obtaining the approximate solution, we investigate the thermodynamic properties of the black hole by analysing important quantities such as the Hawking temperature $T$ and entropy $S$ as functions of the event horizon radius $r_h$.
In addition, we study the specific heat $C_p$ and the Gibbs free energy in order to examine the local and global stability of the black hole. Similar thermodynamic analyses of black holes have also been carried out in other modified gravity theories, including $f(R)$ gravity~\cite{I25,I26,I27}, $f(G)$ gravity~\cite{I28,I29}, $f(T)$ gravity~\cite{I30,I31,I32},$F(R)$-ModMax Theory \cite{panah2024analytic} and $f(P)$ gravity~\cite{aniUD}, among others.
\par
The structure of this paper is organised as follows. In Section \eqref{sec2}, we present the field equations of Hoyle--Narlikar gravity by considering the electric charge as the source. We then obtain an approximate black hole metric solution up to second order in the parameter $\alpha$.In Section \eqref{sec3}, we investigate the thermodynamic properties of the obtained solution by computing important quantities such as the Hawking temperature and entropy. In Section \eqref{sec4}, we calculate the specific heat and Gibbs free energy in order to analyze the local and global thermodynamic stability of the black hole.
\section{BLACK HOLE SOLUTIONS IN  Hoyle–Narlikar GRAVITY}\label{sec2}
In this work, we carry out our investigation in the context of the Hoyle--Narlikar theory of gravitation, a framework in which the conventional Einstein field equations are generalised through the inclusion of a creation-field contribution. Consequently, the gravitational field equations are modified, and the complete system of equations takes the following form \cite{I6,I7,I8}.
\begin{equation}\label{e1}
    R_{\mu \nu} - \frac{1}{2} R g_{\mu \nu} = 8\pi \left[ T^{(m)}_{\mu \nu} + T^{(C)}_{\mu \nu} \right]
\end{equation}
In this expression, $R$ represents the Ricci scalar, $R_{\mu\nu}$ denotes the Ricci curvature tensor, and $g_{\mu\nu}$ is the space--time metric tensor. In addition, $T_{\mu\nu}^{(m)}$ and $T_{\mu\nu}^{(C)}$ refer to the energy--momentum tensors associated with the charged matter distribution and the creation field, respectively.
\begin{equation}\label{e2}
T^{(C)}_{\mu\nu} = -\tilde{a} \left( C_{\mu} C_{\nu} - \frac{1}{2} C_{i} C^{i} \, g_{\mu\nu} \right)
\end{equation}

\begin{equation}\label{e3}
    T^{(m)}_{\mu \nu}= \frac{1}{4\pi} \left( F_{\nu\alpha} F_{\mu}^{\hspace{0.1cm}\alpha}- \frac{1}{4} g_{\mu\nu} F^{\alpha\beta} F_{\alpha\beta} \right)
\end{equation}

Where $F=F_{\mu\nu} F^{\mu\nu}$ and \( F_{\mu\nu} = \partial_\mu A_\nu - \partial_\nu A_\mu \), where \( A_\mu \) is the gauge potential \cite{Sec2.1,Sec2.2,Sec2.3,Sec2.4,Sec2.5}.The Maxwell equation reads as 
\begin{equation}\label{e4}
    \partial_{\nu}\Big( \sqrt{-g}F^{\mu \nu} \Big)=0
\end{equation}
Solving the above equation, we obtain $F_{tr}=\frac{Q}{r^2}$.Where Q is the integration constant corresponding to the charge.
The scalar function $C$ represents the creation field, and its corresponding four-vector is defined by
\[
C_{\mu} = \frac{\partial C}{\partial x^{\mu}}\,,
\]
where $\tilde{a} > 0$ denotes the associated coupling constant.We now proceed to obtain the black hole solution in the context of Hoyle--Narlikar gravity. Now, we consider a static and spherically symmetric space-time background, described by the following line element:
\begin{equation}\label{e5}
    ds^2 = -\psi(r) \, dt^2 + \frac{dr^2}{\psi(r)} + r^2 \left( d\theta^2 + \sin^2\theta \, d\phi^2 \right),
\end{equation}
Here, $\psi(r)$ represents an arbitrary function to be determined from the modified field equations. Moreover, we assume that the creation field $C$ depends only on the radial coordinate $r$. In particular, we consider a linear radial dependence of the form
\[
C(r) \propto r.
\]
Substituting Eq.~\eqref{e5}  into Eq.~\eqref{e1}, we obtain the charged black hole metric \eqref{e6} up to second-order corrections in the parameter $\alpha$ within the framework of Hoyle--Narlikar gravity.
\begin{widetext}
\begin{equation}\label{e6}
    \psi(r)=
\left(1 + \frac{Q^{2}}{r^{2}} +C_{1}\right)
+ \left(4\pi Q^{2} - 4\pi M r + \frac{4\pi r^{2}}{3}\right)\alpha
+ \left(
\frac{16}{3}\pi^{2} Q^{2} r^{2}
- 4\pi^{2} M r^{3}
+ \frac{16\pi^{2} r^{4}}{15}
\right)\alpha^{2}
\end{equation}
\end{widetext}
\begin{figure*}[htbp]
    \centering
    \begin{subfigure}{0.49\textwidth}
    \centering
    \includegraphics[width=1\linewidth]{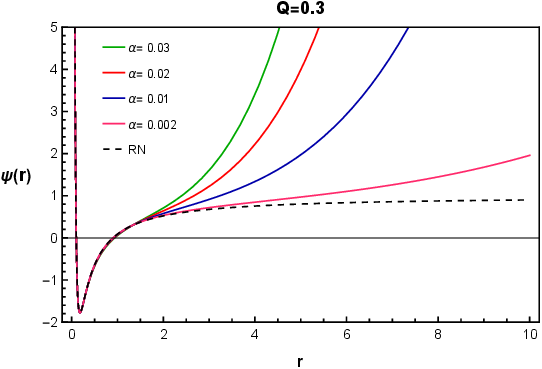}
    \caption{Q=0.3} 
    \label{f1a}
    \end{subfigure}
    \hfill
    \begin{subfigure}{0.49\textwidth}
    \centering
    \includegraphics[width=1\linewidth]{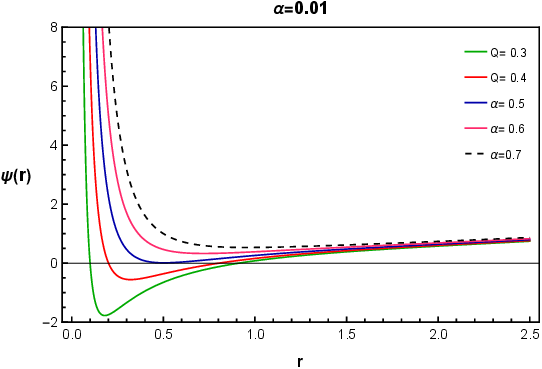}
    \caption{$\alpha=0.01$}
    \label{f1b}
    \end{subfigure}
    \hfill
      \caption{We plot the $\psi(r)$ vs $r$ for $M=1/2$.
} \label{f1}
    \end{figure*}
Where \(C_{1} = -2M\) is an integration constant associated with the black hole mass, and \(\alpha\) is the parameter corresponding to Hoyle--Narlikar gravity. In the limit \(\alpha \to 0\), the solution reduces to the Reissner--Nordström (RN) metric. In this context, our aim is to determine the real roots of the obtained metric function given in Eq.~\eqref{e6}, as these roots provide crucial information regarding the location of the horizons (namely, the inner and outer horizons) of the solution. Black holes are characterised by the presence of a curvature singularity located at $r = 0$. However, it is also possible to obtain solutions without an event horizon; such configurations are referred to as naked singularities. To determine the roots, it is convenient to analyse the metric function directly. Since the metric function is a fourth-order polynomial in $r$, obtaining an exact analytical solution is rather complicated. Therefore, we plot the metric function as a function of $r$ in Fig.~\eqref{f1} in order to identify its real roots. As illustrated in Fig.~\eqref{f1}, two distinct cases are considered. In the first case, the charge $Q$ is kept fixed while the parameter $\alpha$ is varied. In the second case, the parameter $\alpha$ is fixed, and the charge $Q$ is varied.
In the first case, two real roots are obtained, corresponding to the inner (Cauchy) horizon and the outer (event) horizon. In the second case, three possibilities may arise:
(i) two distinct real roots exist, indicating the presence of both an inner (Cauchy) horizon and an outer (event) horizon,
(ii) one real root exists, corresponding to the extremal case in which the two horizons coincide and
(iii) no real root exists, implying the absence of a horizon and, consequently, the formation of a naked singularity.
\par
Now, we examine the effects of the parameter $\alpha$, which corresponds to Hoyle--Narlikar gravity, and the electric charge $Q$ on the black hole metric. In Fig.~\eqref{f1a}, the charge $Q$ is kept fixed while the parameter $\alpha$ is varied. We observe (See Fig.\eqref{f1a})that the black hole metric exhibits two horizons. When $\alpha$ is taken to be very small, the solution approaches the Reissner--Nordström (RN) metric, represented by the black dotted line. However, as $\alpha$ increases, the metric function deviates significantly from the RN case, and the corresponding curves separate more rapidly. From Fig.~\eqref{f1b}, we observe that the effect of the charge indicates that as $Q$ increases, the number of roots decreases. In other words, for sufficiently large values of $Q$, the black hole in Hoyle--Narlikar gravity no longer possesses an event horizon, leading to the formation of a naked singularity.
\section{Thermodynamics}\label{sec3}
This section focuses on the thermodynamic \cite{Sec3.1,Sec3.2, Sec3.3,Sec3.4,Sec3.5,Sec3.6,Sec3.7,Sec3.8} properties of the charged black hole in Hoyle--Narlikar gravity.The first step involves the computation of the Hawking temperature of the black hole. The Hawking temperature \cite{Sec3.9} can be determined from the following equations:
\begin{equation}\label{e7}
T=\frac{k}{2 \pi}
\end{equation}
Where $k$ is the surface gravity of the Black hole. Which is given by
\begin{equation}\label{e8}
      \left.k=\frac{\psi'(r_{h})}{2} \right\vert_{r=r_h}
\end{equation}
Where $r_{h}$ is the radius of the event horizon. The black hole mass can be obtained from the condition $\psi(r_h)=0$, where $r_h$ denotes the horizon radius. Therefore, for the metric function given in Eq.~\eqref{e6}, and by using Eqs.~\eqref{e7} and \eqref{e8}, the mass and temperature can be determined as follows:
\begin{equation}\label{e9}
    M=\frac{Q^{2} + r_h^{2}}{2 r_h}
+
\pi \left(Q^{2} r_h - \frac{r_h^{3}}{3}\right)\alpha
+
\frac{\pi^{2} r_h^{3}}{15}
\left(-5 Q^{2} + 3 r_h^{2}\right)\alpha^{2}
\end{equation}
\begin{equation}\label{e10}
T=\frac{-Q^{2} + r_h^{2}}{4\pi r_h^{3}}
+
\frac{2}{15}\,\pi^{2} r_h^{3}
\left(-25 Q^{2} + 7 r_h^{2}\right)
\alpha^{3}
\end{equation}
From Eq.~\eqref{e10}, it is evident that the Hawking temperature of the black hole in Hoyle--Narlikar gravity is explicitly governed by both the electric charge $Q$ and the parameter $\alpha$. This indicates that the thermal behaviour of the black hole is significantly influenced by the combined effects of the charge and the modified gravitational coupling. Variations in either $Q$ or $\alpha$ alter the horizon structure, which in turn modifies the surface gravity and, consequently, the Hawking temperature.
Within the framework of classical thermodynamics, only positive values of temperature correspond to physically meaningful equilibrium states. A positive Hawking temperature signifies that the black hole radiates energy in a thermodynamically consistent manner. In contrast, negative temperature values are generally regarded as non-physical in this context, as they would imply an ill-defined or unstable thermodynamic configuration. 
\begin{figure*}[htbp]
    \centering
    \begin{subfigure}{0.49\textwidth}
    \centering
    \includegraphics[width=1\linewidth]{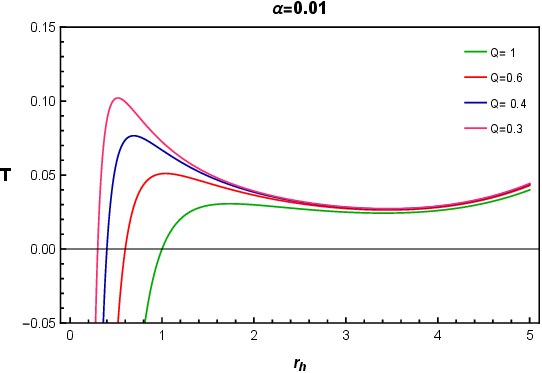}
    \caption{$\alpha$=0.01} 
    \label{f2a}
    \end{subfigure}
    \hfill
    \begin{subfigure}{0.49\textwidth}
    \centering
    \includegraphics[width=1\linewidth]{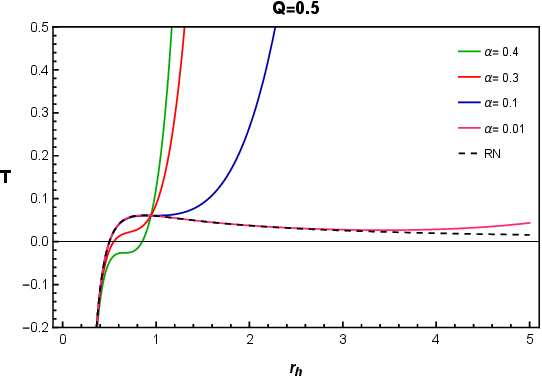}
    \caption{$Q=0.5$}
    \label{f2b}
    \end{subfigure}
    \hfill
      \caption{The Hawking temperature $T$ vs $r_{h}$ for different values of parameters.
} \label{f2}
    \end{figure*}
    Fig.~\eqref{f2} illustrates the Hawking temperature of the black hole in Hoyle--Narlikar gravity for a fixed value of the parameter $\alpha$, while varying the charge $Q$. From the figure, we observe that for small horizon radii, the temperature becomes negative, indicating that the black hole is thermodynamically unstable in this region. As the horizon radius increases, the temperature rises and reaches a maximum value. Beyond this point, the temperature decreases and then increases again with further increase in the horizon radius. This behaviour suggests the presence of different thermodynamic phases. In particular, for sufficiently large horizon radii, the temperature remains positive, implying that the black hole becomes thermodynamically stable in the large-horizon regime.
\par
Fig.~\eqref{f2b} illustrates the Hawking temperature of the black hole in Hoyle--Narlikar gravity for a fixed value of the charge $Q$, while varying the parameter $\alpha$. From the figure, we observe that for small horizon radii, the temperature becomes negative, indicating that the black hole is thermodynamically unstable in this region.
As the horizon radius increases, the temperature rises and eventually remains positive. In particular, for sufficiently large horizon radii, the positive temperature indicates that the black hole becomes thermodynamically stable in the large-horizon regime.

We also observe that for a small value of the parameter, $\alpha = 0.01$, the corresponding curve nearly coincides with the Hawking temperature of the Reissner--Nordström (RN) metric. However, for larger horizon radii, the curve begins to increase more rapidly, clearly demonstrating the influence of the parameter $\alpha$.
\par
On the other hand, the entropy can be calculated using the following formula:
\begin{equation}\label{e11}
S = \int \frac{1}{T}\, dM
\end{equation}
After solving Eq.~\eqref{e11}, we obtain the entropy of the charged black hole in Hoyle--Narlikar gravity, which is given by Eq.~\eqref{e12}.
\begin{equation}\label{e12}
S=\pi r_h^{2}   + \frac{\pi}{4\alpha}
-\pi^{2} r_h^{4}\alpha
+\frac{2}{3}\pi^{3} r_h^{6}\alpha^{2}
\end{equation}
From Eq.~\eqref{e12}, we observe that the entropy does not depend on the charge $Q$; rather, it depends only on the parameter $\alpha$.
\begin{figure}
    \centering
    \includegraphics[width=1\linewidth]{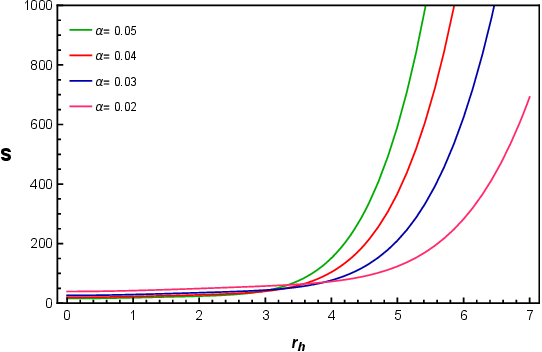}
    \caption{S vs $r_{h}$}
    \label{f3}
\end{figure}
From Fig.~\eqref{f3}, it is evident that the entropy increases monotonically with the horizon radius. This behaviour is consistent with the general thermodynamic expectation that a larger black hole possesses a greater number of accessible microstates. Since the horizon radius is directly related to the geometric size of the black hole, an increase in $r_h$ effectively enlarges the horizon area, which in turn leads to an increase in entropy.

This result also aligns with the well-known area law of black hole thermodynamics, according to which the entropy is proportional to the horizon area. Therefore, as the horizon radius grows, the entropy increases accordingly, indicating a thermodynamically consistent behaviour in the framework of Hoyle--Narlikar gravity.
\section{Thermal Stability}\label{sec4}
The thermodynamic stability of the black hole can be assessed by analysing its response to small thermal fluctuations. In this context, we explore how the parameters of Hoyle--Narlikar gravity influence the stability behaviour. To achieve this, we evaluate key thermodynamic quantities, namely the heat capacity and the Gibbs free energy, which provide insight into phase transitions and equilibrium properties of the system.
\subsection{Specific Heat}
In the canonical ensemble, the thermal stability of a system can be studied by analysing its heat capacity. The sign of the heat capacity determines whether the system is stable or unstable. A positive heat capacity indicates that the system is thermally stable, while a negative heat capacity implies instability.

Therefore, to study the local stability of black holes in Hoyle--Narlikar gravity, we calculate the heat capacity. The heat capacity is given by:
\begin{equation}\label{e13}
C_p = T \left( \frac{\partial S}{\partial T} \right)
\end{equation}
By substituting Eqs.~\eqref{e10} and \eqref{e12} into Eq.~\eqref{e13}, the specific heat capacity can be written as follows:
\begin{widetext}
\begin{equation}\label{e14}
C_p= 
\frac{
\left( 2\pi r_h - 4\pi^2 r_h^3 \alpha + 4\pi^3 r_h^5 \alpha^2 \right)
\left( \frac{-Q^2 + r_h^2}{4\pi r_h^3}
+ \frac{2}{15}\pi^2 r_h^3 (-25 Q^2 + 7 r_h^2)\alpha^3 \right)
}{
\frac{1}{2\pi r_h^2}
- \frac{3(-Q^2 + r_h^2)}{4\pi r_h^4}
+ \frac{28}{15}\pi^2 r_h^4 \alpha^3
+ \frac{2}{5}\pi^2 r_h^2 (-25 Q^2 + 7 r_h^2)\alpha^3
}
\end{equation}
\end{widetext}
\begin{figure*}[htbp]
    \centering
    \begin{subfigure}{0.49\textwidth}
    \centering
    \includegraphics[width=1\linewidth]{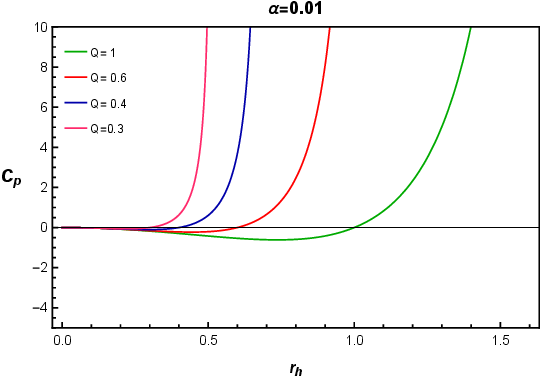}
    \caption{$\alpha$=0.01} 
    \label{f4a}
    \end{subfigure}
    \hfill
    \begin{subfigure}{0.49\textwidth}
    \centering
    \includegraphics[width=1\linewidth]{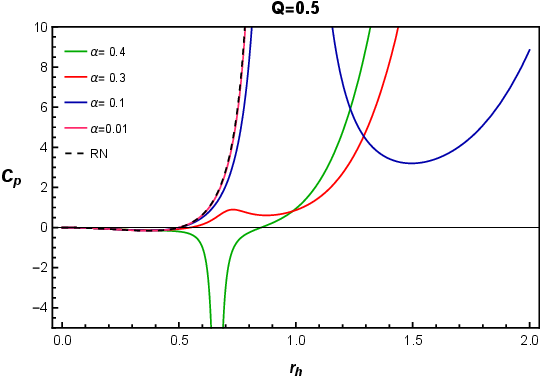}
    \caption{$Q=0.5$}
    \label{f4b}
    \end{subfigure}
    \hfill
      \caption{$C_{p}$ vs $r_h$} \label{f4}
    \end{figure*}
In black hole thermodynamics, the zeros of the heat capacity $C$ are often interpreted as boundary points separating non-physical and physical configurations. At these points, the heat capacity changes its sign, indicating a transition between thermodynamically unstable and stable regions.
Moreover, divergences of the heat capacity are commonly regarded as signals of phase transition critical points. Such divergences suggest the presence of second-order phase transitions in the black hole system. Therefore, by analysing both the zeros and the divergences of the heat capacity, one can determine the physical limitation points as well as the phase transition critical points of black holes.
From Fig.~\eqref{f4a}, we observe that when the parameter $\alpha$ is fixed, and the charge $Q$ is varied, the heat capacity is negative for small values of the horizon radius, indicating thermodynamic instability in this region. As the horizon radius increases, the black hole undergoes a phase transition, after which the heat capacity becomes positive, signalling a stable thermodynamic phase.

We also note that increasing the charge $Q$ shifts the phase transition point toward larger values of the horizon radius. This implies that a higher charge delays the onset of thermodynamic stability.
From Fig.~\eqref{f4b}, we observe that when the charge $Q$ is fixed and the parameter $\alpha$ is varied, the heat capacity becomes positive for large values of the horizon radius. This indicates that the charged black hole in Hoyle--Narlikar gravity is thermodynamically stable in the large-horizon regime. 

Additionally, the black dotted line represents the corresponding result for the Reissner--Nordström (RN) black hole metric.
\subsection{Gibbs free energy}
To gain deeper insight into the phase structure and stability of the black hole, we compute the Gibbs free energy. It is important to note that, in the extended phase space framework, the black hole mass is interpreted as the enthalpy rather than the internal energy. Therefore, the Gibbs free energy can be obtained from the relation:
\begin{equation}\label{e15}
G = M - T S
\end{equation}
By substituting the expressions for the mass, temperature, and entropy into Eq.~\eqref{e15}, the Gibbs free energy can be written as follows:
\begin{widetext}
\begin{equation}\label{e16}
G =
\frac{Q^2 - r_h^2}{16 r_h^3 \alpha}
+ \frac{3Q^2 + r_h^2}{4 r_h}
+ \frac{1}{12} \left( 9\pi Q^2 r_h - \pi r_h^3 \right)\alpha
+ \frac{1}{30} \left( -5\pi^2 Q^2 r_h^3 + 25\pi^3 Q^2 r_h^3
+ \pi^2 r_h^5 - 7\pi^3 r_h^5 \right)\alpha^2
\end{equation}
\end{widetext}
In the canonical ensemble, the global thermodynamic stability of a system is characterised by the sign of the Gibbs free energy. A negative Gibbs free energy indicates that the system is globally stable. 
Therefore, to analyse the global stability of black holes in Hoyle--Narlikar gravity, we evaluate the behaviour of the Gibbs free energy [Eq.\eqref{e16}].
From Fig.~\eqref{e5}, we observe that the Gibbs free energy becomes negative for large values of the horizon radius. Therefore, the black hole in Hoyle--Narlikar gravity is globally thermodynamically stable in the large-horizon regime.
As a result, we find that large black holes in Hoyle--Narlikar gravity satisfy both the local and global stability conditions in the large-horizon regime. This conclusion is obtained by comparing the regions of local stability (from Fig.~\eqref{f4}) with those of global stability (from Fig.~\eqref{e5}), where both conditions are simultaneously fulfilled.
\begin{figure*}[htbp]
    \centering
    \begin{subfigure}{0.49\textwidth}
    \centering
    \includegraphics[width=1\linewidth]{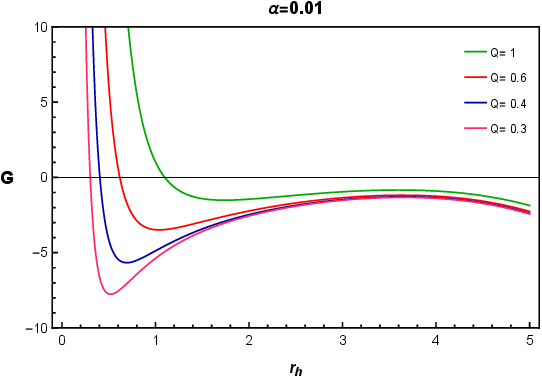}
    \caption{$\alpha$=0.01} 
    \label{f5a}
    \end{subfigure}
    \hfill
    \begin{subfigure}{0.49\textwidth}
    \centering
    \includegraphics[width=1\linewidth]{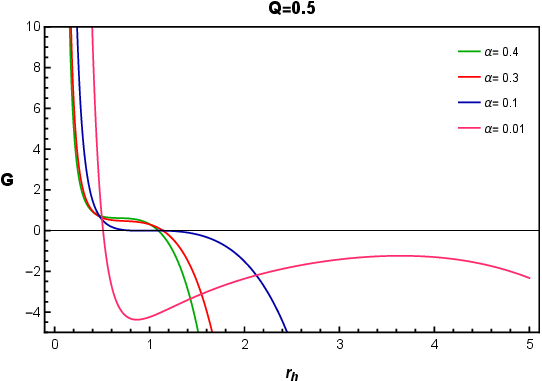}
    \caption{$Q=0.5$}
    \label{f5b}
    \end{subfigure}
    \hfill
      \caption{$G$ vs $r_{h}$} \label{f5}
    \end{figure*}
\section{SUMMARY AND CONCLUSIONS}\label{sec5}
In this work, we construct a new black hole solution within the framework of Hoyle--Narlikar gravity, where the electric charge acts as the source surrounding the black hole. We also show that the obtained metric reduces to the Reissner--Nordström (RN) metric in the limit $\alpha \to 0$. Subsequently, we compute the Hawking temperature, entropy, specific heat, and Gibbs free energy in order to analyse the thermodynamic stability of the black hole.
\par
(i) First, we have shown that the theory admits a charged black hole solution that is static and spherically symmetric, characterised by a single metric function $\psi(r)$ given in Eq.~\eqref{e6}. Notably, this solution can be regarded as a generalisation of the Reissner--Nordström (RN) black hole. Furthermore, as illustrated in Fig.~\eqref{e1}, the metric structure within the framework of Hoyle--Narlikar gravity admits three possible cases: two horizons,one horizon, or no horizon, depending on the values of the parameter and charge Q.
\par
(ii) After deriving the metric function, we proceed to investigate the thermodynamic properties of the charged black hole in Hoyle--Narlikar gravity by computing the Hawking temperature and entropy. The behaviour of the Hawking temperature reveals that, although it may become negative for a small horizon radius, it turns positive as the horizon radius increases. This indicates that large black holes correspond to physically meaningful and thermodynamically stable configurations. In addition, the entropy is found to increase monotonically with the horizon radius $r_h$. This behaviour is consistent with the fundamental principles of black hole thermodynamics, where entropy is directly related to the size (or area) of the event horizon. Therefore, larger black holes possess greater entropy, reflecting a higher number of microscopic degrees of freedom and a more stable thermodynamic state.
\par
(iii) Finally, we have computed the specific heat and the Gibbs free energy to analyse the local and global stability of the charged black hole in Hoyle--Narlikar gravity. For large values of the horizon radius, the specific heat is positive, while the Gibbs free energy becomes negative. These behaviours indicate that the black hole is both locally and globally stable in this regime.
Therefore, we conclude that large black holes in Hoyle--Narlikar gravity satisfy the conditions for both local and global thermodynamic stability in the large-horizon limit. This conclusion is drawn by comparing the regions of local stability shown in Fig.~\eqref{f4} with the regions of global stability displayed in Fig.~\eqref{f5}, where both stability criteria are simultaneously satisfied.
\section{Acknowledgements}
AG is thankful to IIEST, Shibpur, India, for providing the Institute
Fellowship (SRF).
\section{Data Availability Statement}
 This manuscript has no associated data.
[Author’s comment: In the present study, no datasets are generated or
analyzed]

\bibliographystyle{naturemag}
\bibliography{bibliography}

@article{I1,
  title={Introduction to modified gravity and gravitational alternative for dark energy},
  author={Nojiri, Shin'Ichi and Odintsov, Sergei D},
  journal={International Journal of Geometric Methods in Modern Physics},
  volume={4},
  number={01},
  pages={115--145},
  year={2007},
  publisher={World Scientific}
}

@article{I2,
title = {Modified gravity and cosmology},
journal = {Physics Reports},
volume = {513},
number = {1},
pages = {1-189},
year = {2012},
note = {Modified Gravity and Cosmology},
issn = {0370-1573},
doi = {https://doi.org/10.1016/j.physrep.2012.01.001},
url = {https://www.sciencedirect.com/science/article/pii/S0370157312000105},
author = {Timothy Clifton and Pedro G. Ferreira and Antonio Padilla and Constantinos Skordis},

}

@article{I3,
title = {Unified cosmic history in modified gravity: From F(R) theory to Lorentz non-invariant models},
journal = {Physics Reports},
volume = {505},
number = {2},
pages = {59-144},
year = {2011},
issn = {0370-1573},
doi = {https://doi.org/10.1016/j.physrep.2011.04.001},
url = {https://www.sciencedirect.com/science/article/pii/S0370157311001335},
author = {Shin’ichi Nojiri and Sergei D. Odintsov},}

@article{I4,
  title={Dynamics of dark energy},
  author={Copeland, Edmund J and Sami, Mohammad and Tsujikawa, Shinji},
  journal={International Journal of Modern Physics D},
  volume={15},
  number={11},
  pages={1753--1935},
  year={2006},
  publisher={World Scientific}
}

@article{I5,
  title={f (T) teleparallel gravity and cosmology},
  author={Cai, Yi-Fu and Capozziello, Salvatore and De Laurentis, Mariafelicia and Saridakis, Emmanuel N},
  journal={Reports on Progress in Physics},
  volume={79},
  number={10},
  pages={106901},
  year={2016},
  publisher={IOP Publishing}
}

@article{I6,
  title={On the avoidance of singularities in C-field cosmology},
  author={Hoyle, Fred and Narlikar, Jayant Vishnu},
  journal={Proceedings of the Royal Society of London. Series A. Mathematical and Physical Sciences},
  volume={278},
  number={1375},
  pages={465--478},
  year={1964},
  publisher={The Royal Society London}
}

@article{I7,
  title={The C-field as a direct particle field},
  author={Hoyle, Fred and Narlikar, JV},
  journal={Proceedings of the Royal Society of London. Series A. Mathematical and Physical Sciences},
  volume={282},
  number={1389},
  pages={178--183},
  year={1964},
  publisher={The Royal Society London}
}

@article{I8,
  title={A new theory of gravitation},
  author={Hoyle, Fred and Narlikar, Jayant V},
  journal={Proceedings of the Royal Society of London. Series A. Mathematical and Physical Sciences},
  volume={282},
  number={1389},
  pages={191--207},
  year={1964},
  publisher={The Royal Society London}
}

@article{I9,
  title={Singularity and matter creation in cosmological models},
  author={Narlikar, JV},
  journal={Nature Physical Science},
  volume={242},
  number={122},
  pages={135--136},
  year={1973},
  publisher={Nature Publishing Group UK London}
}

@article{I10,
  title={Creation-field cosmology: A possible solution to singularity, horizon, and flatness problems},
  author={Narlikar, JV and Padmanabhan, T},
  journal={Physical Review D},
  volume={32},
  number={8},
  pages={1928},
  year={1985},
  publisher={APS}
}

@article{I11,
  title={On the Hoyle-Narlikar theory of gravitation},
  author={Hawking, Stephen William},
  journal={Proceedings of the Royal Society of London. Series A. Mathematical and Physical Sciences},
  volume={286},
  number={1406},
  pages={313--319},
  year={1965},
  publisher={The Royal Society London}
}

@article{I12,
  title={A new model for the expanding universe},
  author={Hoyle, Fred},
  journal={Monthly Notices of the Royal Astronomical Society, Vol. 108, p. 372},
  volume={108},
  pages={372},
  year={1948}
}

@article{I13,
  title={A Generalized Hoyle--Narlikar Particle Theory},
  author={McIntosh, CBG},
  journal={Nature},
  volume={226},
  number={5243},
  pages={339--340},
  year={1970},
  publisher={Nature Publishing Group UK London}
}

@article{I14,
  title={Hoyle-Narlikar theory of gravitation},
  author={Davies, PCW},
  journal={Nature},
  volume={228},
  number={5268},
  pages={270--271},
  year={1970},
  publisher={Nature Publishing Group UK London}
}

@article{I15,
  title={A conformal theory of gravitation},
  author={Hoyle, Fred and Narlikar, Jayant Vishnu},
  journal={Proceedings of the Royal Society of London. Series A. Mathematical and Physical Sciences},
  volume={294},
  number={1437},
  pages={138--148},
  year={1966},
  publisher={The Royal Society London}
}

@book{I16,
  title={A different approach to cosmology: from a static universe through the big bang towards reality},
  author={Hoyle, Fred and Burbidge, Geoffrey and Narlikar, Jayant Vishnu},
  year={2000},
  publisher={Cambridge University Press}
}

@article{I17,
  title={Primordial element formation, primordial magnetic fields, and the isotropy of the universe},
  author={Thorne, Kip S},
  journal={Astrophysical Journal, vol. 148, p. 51},
  volume={148},
  pages={51},
  year={1967}
}

@article{I18,
  title={On the Hubble constant and the cosmological constant},
  author={Hoyle, F and Burbidge, G and Narlikar, JV},
  journal={Monthly Notices of the Royal Astronomical Society},
  volume={286},
  number={1},
  pages={173--182},
  year={1997},
  publisher={Blackwell Science Ltd Oxford, UK}
}

@article{I1.1,
  title={Reconstruction of dark energy and expansion dynamics using Gaussian processes},
  author={Seikel, Marina and Clarkson, Chris and Smith, Mathew},
  journal={Journal of Cosmology and Astroparticle Physics},
  volume={2012},
  number={06},
  pages={036},
  year={2012},
  publisher={IOP Publishing}
}

@article{I1.2,
  title = {Direct Reconstruction of Dark Energy},
  author = {Clarkson, Chris and Zunckel, Caroline},
  journal = {Phys. Rev. Lett.},
  volume = {104},
  issue = {21},
  pages = {211301},
  numpages = {4},
  year = {2010},
  month = {May},
  publisher = {American Physical Society},
  doi = {10.1103/PhysRevLett.104.211301},
  url = {https://link.aps.org/doi/10.1103/PhysRevLett.104.211301}
}

@article{I1.3,
  title={Reconstruction of dark energy and equilibrium thermodynamics in Brans-Dicke theory},
  author={Liu, Xian-Ming and Liu, Wen-Biao},
  journal={Astrophysics and Space Science},
  volume={334},
  pages={203--207},
  year={2011},
  publisher={Springer}
}

@article{I1.4,
  title={A theory of dark matter},
  author={Arkani-Hamed, Nima and Finkbeiner, Douglas P and Slatyer, Tracy R and Weiner, Neal},
  journal={Physical Review D—Particles, Fields, Gravitation, and Cosmology},
  volume={79},
  number={1},
  pages={015014},
  year={2009},
  publisher={APS}
}

@article{I1.5,
  title={Dark energy and dark matter},
  author={Comelli, Daniele and Pietroni, M and Riotto, A},
  journal={Physics Letters B},
  volume={571},
  number={3-4},
  pages={115--120},
  year={2003},
  publisher={Elsevier}
}

@article{I1.6,
  title={History of dark matter},
  author={Bertone, Gianfranco and Hooper, Dan},
  journal={Reviews of Modern Physics},
  volume={90},
  number={4},
  pages={045002},
  year={2018},
  publisher={APS}
}

@article{I25,
  title={Black-hole solutions in F (R) gravity with conformal anomaly},
  author={Hendi, SH and Momeni, D},
  journal={The European Physical Journal C},
  volume={71},
  number={12},
  pages={1823},
  year={2011},
  publisher={Springer}
}

@article{I26,
  title={Charged accelerating black hole in f (R) gravity},
  author={Zhang, Ming and Mann, Robert B},
  journal={Physical Review D},
  volume={100},
  number={8},
  pages={084061},
  year={2019},
  publisher={APS}
}

@article{I27,
  title={Gravitational perturbations of a Kerr black hole in f (R) gravity},
  author={Suvorov, Arthur George},
  journal={Physical Review D},
  volume={99},
  number={12},
  pages={124026},
  year={2019},
  publisher={APS}
}

@article{I28,
  title={Regular black holes in f (G) gravity},
  author={de S. Silva, Marcos V and Rodrigues, Manuel E},
  journal={The European Physical Journal C},
  volume={78},
  pages={1--18},
  year={2018},
  publisher={Springer}
}

@article{I29,
  title={Regular multihorizon black holes in f (G) gravity with nonlinear electrodynamics},
  author={Rodrigues, Manuel E and Silva, Marcos V de S},
  journal={Physical Review D},
  volume={99},
  number={12},
  pages={124010},
  year={2019},
  publisher={APS}
}

@article{I30,
  title={Violation of the first law of black hole thermodynamics in f (T) gravity},
  author={Miao, Rong-Xin and Li, Miao and Miao, Yan-Gang},
  journal={Journal of Cosmology and Astroparticle Physics},
  volume={2011},
  number={11},
  pages={033},
  year={2011},
  publisher={IOP Publishing}
}

@article{I31,
  title={Static spherically symmetric black holes in weak f (T)-gravity},
  author={Pfeifer, Christian and Schuster, Sebastian},
  journal={Universe},
  volume={7},
  number={5},
  pages={153},
  year={2021},
  publisher={MDPI}
}

@article{I32,
  title={Exact charged black-hole solutions in D-dimensional f (T) gravity: torsion vs curvature analysis},
  author={Capozziello, Salvatore and Gonzalez, PA and Saridakis, Emmanuel N and Vasquez, Yerko},
  journal={Journal of High Energy Physics},
  volume={2013},
  number={2},
  pages={1--25},
  year={2013},
  publisher={Springer}
}

@article{aniUD,
  title={New Black Hole Solutions in f (P) Gravity and their Thermodynamic Nature},
  author={Ghosh, Aniruddha and Debnath, Ujjal},
  journal={Physics Letters B},
  pages={139305},
  year={2025},
  publisher={Elsevier}
}

@article{panah2024analytic,
  title={Analytic electrically charged black holes in F (R)-ModMax theory},
  author={Panah, Behzad Eslam},
  journal={Progress of Theoretical and Experimental Physics},
  volume={2024},
  number={2},
  pages={023E01},
  year={2024},
  publisher={Oxford University Press}
}

@article{capozziello2002curvature,
  title={Curvature quintessence},
  author={Capozziello, Salvatore},
  journal={International Journal of Modern Physics D},
  volume={11},
  number={04},
  pages={483--491},
  year={2002},
  publisher={World Scientific}
}

@article{carroll2004cosmic,
  title={Is cosmic speed-up due to new gravitational physics?},
  author={Carroll, Sean M and Duvvuri, Vikram and Trodden, Mark and Turner, Michael S},
  journal={Physical Review D},
  volume={70},
  number={4},
  pages={043528},
  year={2004},
  publisher={APS}
}

@article{sotiriou2006f,
  title={f (R) gravity and scalar--tensor theory},
  author={Sotiriou, Thomas P},
  journal={Classical and Quantum Gravity},
  volume={23},
  number={17},
  pages={5117--5128},
  year={2006}
}

@article{hu2007models,
  title={Models of f (R) cosmic acceleration that evade solar system tests},
  author={Hu, Wayne and Sawicki, Ignacy},
  journal={Physical Review D—Particles, Fields, Gravitation, and Cosmology},
  volume={76},
  number={6},
  pages={064004},
  year={2007},
  publisher={APS}
}

@article{arkani2009theory,
  title={A theory of dark matter},
  author={Arkani-Hamed, Nima and Finkbeiner, Douglas P and Slatyer, Tracy R and Weiner, Neal},
  journal={Physical Review D—Particles, Fields, Gravitation, and Cosmology},
  volume={79},
  number={1},
  pages={015014},
  year={2009},
  publisher={APS}
}

@article{bertone2018history,
  title={History of dark matter},
  author={Bertone, Gianfranco and Hooper, Dan},
  journal={Reviews of Modern Physics},
  volume={90},
  number={4},
  pages={045002},
  year={2018},
  publisher={APS}
}

@article{cirelli2026dark,
  title={Dark matter},
  author={Cirelli, Marco and Strumia, Alessandro and Zupan, Jure},
  journal={SciPost Physics Reviews},
  pages={001},
  year={2026}
}

@article{copeland2006dynamics,
  title={Dynamics of dark energy},
  author={Copeland, Edmund J and Sami, Mohammad and Tsujikawa, Shinji},
  journal={International Journal of Modern Physics D},
  volume={15},
  number={11},
  pages={1753--1935},
  year={2006},
  publisher={World Scientific}
}

@article{li2011dark,
  title={Dark energy},
  author={Li, Miao and Li, Xiao-Dong and Wang, Shuang and Wang, Yi},
  journal={Communications in theoretical physics},
  volume={56},
  number={3},
  pages={525--604},
  year={2011}
}

@article{frieman2008dark,
  title={Dark energy and the accelerating universe},
  author={Frieman, Joshua A and Turner, Michael S and Huterer, Dragan},
  journal={Annu. Rev. Astron. Astrophys.},
  volume={46},
  number={1},
  pages={385--432},
  year={2008},
  publisher={Annual Reviews}
}

@article{comelli2003dark,
  title={Dark energy and dark matter},
  author={Comelli, Daniele and Pietroni, M and Riotto, A},
  journal={Physics Letters B},
  volume={571},
  number={3-4},
  pages={115--120},
  year={2003},
  publisher={Elsevier}
}

@article{Sec2.1,
  title={Charged spherically symmetric black holes in f (R) gravity and their stability analysis},
  author={Nashed, Gamal GL and Capozziello, Salvatore},
  journal={Physical Review D},
  volume={99},
  number={10},
  pages={104018},
  year={2019},
  publisher={APS}
}

@article{Sec2.2,
  title={Higher-dimensional charged black hole solutions with a nonlinear electrodynamics source},
  author={Hassaine, Mokhtar and Martinez, Cristian},
  journal={Classical and Quantum Gravity},
  volume={25},
  number={19},
  pages={195023},
  year={2008}
}

@article{Sec2.3,
  title={Thermodynamics of charged black holes with a nonlinear electrodynamics source},
  author={Gonzalez, Hernan A and Hassaine, Mokhtar and Martinez, Cristian},
  journal={Physical Review D—Particles, Fields, Gravitation, and Cosmology},
  volume={80},
  number={10},
  pages={104008},
  year={2009},
  publisher={APS}
}

@article{Sec2.4,
  title={Non-linear charged dS black hole and its thermodynamics and phase transitions},
  author={Nam, Cao H},
  journal={The European Physical Journal C},
  volume={78},
  number={5},
  pages={418},
  year={2018},
  publisher={Springer}
}

@article{Sec2.5,
  title={Charged black holes in Einsteinian cubic gravity and nonuniqueness},
  author={Frassino, Antonia M and Rocha, Jorge V},
  journal={Physical Review D},
  volume={102},
  number={2},
  pages={024035},
  year={2020},
  publisher={APS}
}

@article{Sec3.1,
  title={The four laws of black hole mechanics},
  author={Bardeen, James M and Carter, Brandon and Hawking, Stephen W},
  journal={Communications in mathematical physics},
  volume={31},
  pages={161--170},
  year={1973},
  publisher={Springer}
}

@article{Sec3.2,
  title = {Black Holes and Entropy},
  author = {Bekenstein, Jacob D.},
  journal = {Phys. Rev. D},
  volume = {7},
  issue = {8},
  pages = {2333--2346},
  numpages = {0},
  year = {1973},
  month = {Apr},
  publisher = {American Physical Society},
  doi = {10.1103/PhysRevD.7.2333},
  url = {https://link.aps.org/doi/10.1103/PhysRevD.7.2333}
}

@article{Sec3.3,
  title = {Generalized second law of thermodynamics in black-hole physics},
  author = {Bekenstein, Jacob D.},
  journal = {Phys. Rev. D},
  volume = {9},
  issue = {12},
  pages = {3292--3300},
  numpages = {0},
  year = {1974},
  month = {Jun},
  publisher = {American Physical Society},
  doi = {10.1103/PhysRevD.9.3292},
  url = {https://link.aps.org/doi/10.1103/PhysRevD.9.3292}
}

@article{Sec3.4,
  title = {Some properties of the Noether charge and a proposal for dynamical black hole entropy},
  author = {Iyer, Vivek and Wald, Robert M.},
  journal = {Phys. Rev. D},
  volume = {50},
  issue = {2},
  pages = {846--864},
  numpages = {0},
  year = {1994},
  month = {Jul},
  publisher = {American Physical Society},
  doi = {10.1103/PhysRevD.50.846},
  url = {https://link.aps.org/doi/10.1103/PhysRevD.50.846}
}

@article{Sec3.5,
  title={Particle creation by black holes},
  author={Hawking, Stephen W},
  journal={Communications in mathematical physics},
  volume={43},
  number={3},
  pages={199--220},
  year={1975},
  publisher={Springer}
}

@article{Sec3.6,
  title = {Black hole entropy is the Noether charge},
  author = {Wald, Robert M.},
  journal = {Phys. Rev. D},
  volume = {48},
  issue = {8},
  pages = {R3427--R3431},
  numpages = {0},
  year = {1993},
  month = {Oct},
  publisher = {American Physical Society},
  doi = {10.1103/PhysRevD.48.R3427},
  url = {https://link.aps.org/doi/10.1103/PhysRevD.48.R3427}
}

@article{Sec3.7,
  title = {Some properties of the Noether charge and a proposal for dynamical black hole entropy},
  author = {Iyer, Vivek and Wald, Robert M.},
  journal = {Phys. Rev. D},
  volume = {50},
  issue = {2},
  pages = {846--864},
  numpages = {0},
  year = {1994},
  month = {Jul},
  publisher = {American Physical Society},
  doi = {10.1103/PhysRevD.50.846},
  url = {https://link.aps.org/doi/10.1103/PhysRevD.50.846}
}

@article{Sec3.8,
  title={Charged AdS black hole thermodynamics in Einstein-Gauss-Bonnet gravity under quintessence field: heat engine},
  author={Mondal, Debojyoti and Debnath, Ujjal},
  journal={Nuclear Physics B},
  pages={117247},
  year={2025},
  publisher={Elsevier}
}

@article{Sec3.9,
  title={Particle creation by black holes},
  author={Hawking, Stephen W},
  journal={Communications in mathematical physics},
  volume={43},
  number={3},
  pages={199--220},
  year={1975},
  publisher={Springer}
}
\end{document}